# Survey on biomarkers in human vocalizations

**Aki Ha¨rma**[*1], **Bert den Brinker**[2], **Ulf Grossekatho¨fer**[2], **Okke Ouweltjes**[2],
**Srikanth Nallanthighal**[3], **Sidharth Abrol**[3] and **Vibhu Sharma**[2]

August 8, 2024

## Abstract

Recent years have witnessed an increase in technologies that use speech for the sensing of the health of the talker. This survey paper proposes a general taxonomy of the technologies and a broad overview of current progress and challenges. Vocal biomarkers are often secondary measures that approximate a signal from another sensor or identify an underlying mental, cognitive, or physiological state. Their measurement involves disturbances and uncertainties that may be considered as noise sources, and the biomarkers are coarsely qualified in terms of the various sources of noise involved in their determination. While in some proposed biomarkers the error levels seem high, there are vocal biomarkers where the errors are expected to be low, and thus are more likely to qualify as candidates for adoption in healthcare applications.

**Index Terms**: Pathological speech sensing, speech breathing, respiratory health and fitness, automatic speech recognition

## 1   Introduction

Human language is said to go back in time at least 1.5 million years [1] but animal vocalization naturally has a much longer history [2]. Animal vocalization and human speech can be seen as adaptations of the respiratory system to communication. Vocalization requires a high level of neural control of the muscles in the vocal track consisting of the lungs, trachea, and mouth. Speech, with a communicative message, results from cognitive processes in the brain that reflect all aspects of human life. In contrast, mental, cognitive, and physiological changes can have an impact on vocalization. Vocal biomarkers, acoustic signatures extracted from speech and other vocalizations, are therefore expected to reflect these.

Surveys on the use of voice for health and fitness monitoring have been published, for example, by Fagrehazzi et al. [3] and Ferro et al. [4]. Kraus defines a vocal biomarker as a signature, a feature or a combination of features of the voice that is associated with a clinical outcome and can be used to monitor patients, diagnose a condition, or grade severity or stages of a disease or for drug development [5]. For the current paper, we use a broader definition of vocal biomarker technology as systems that aim at using human vocalization, captured with a microphone, to detect a condition of the talker, or estimate the measurement values of another sensory modality.

Auscultation based on listening to the sounds of the heart, lungs, and intestine, using a stethoscope, for example, has a long history in medicine dating from antiquity to today [6, 7]. In this report, we focus on vocalizations captured from air using a microphone. We leave out, for example, breathing sounds captured using microphones of various kinds and other bodily sounds not produced by air flow passing through the vocal tract. We also leave out ingressive vocalizations [8] and sounds captured by bone conduction, e.g., by in-ear headphones or eyeglasses.

In vocalizations, the main interest is in conversational speech, for example, captured during a telephone call or in using a speech user interface in a smart speaker. However, there are several other types of vocalization that we want to cover. Some of those have been listed in Table 1, which also shows some of the detection targets such as asthma,

---
[*1] DACS, Maastricht University, The Netherlands. [2]Philips Research, Eindhoven, The Netherlands. [3]Philips Research, Bangalore, India

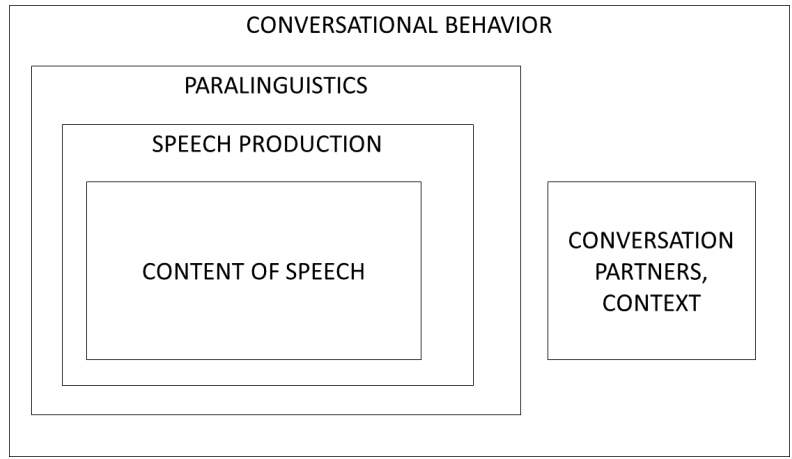

Figure 1: Different aspects of speech communication research

depression, or alcohol intoxication. In estimating measurement values of another sensing modality, our own case is Virtual Respiratory Belt (VRB) processing where we estimate from speech a waveform of a respiratory belt sensor signal of the talker. The opposite cases of sensing speech from other modalities, for example, silent-speech interfaces based on speech movements or BMI [9], or methods focusing on speech reception testing [10] are out of the scope of this review.

| Vocalization | Breathing sounds | Involuntary non-verbal vocalizations | Other non-verbal vocalization | Spontaneous speech | Protocolled speech |
|---|---|---|---|---|---|
| Examples | Normal breathing, wheezing, crackles | Snoring, groaning, grunting, tics | Coughing, sneezing, hiccups, back-channeling | Conversational speech, free speech | Read speech, therapeutic protocols |
| Condition | Asthma, COPD, covid-19 | Apneas, Tourette's syndrome | Infections, COPD, alcohol, depression | Respiratory health, AD, PD, depression | Dementia, arrythmias |

Table 1: Types of vocalizations

Speech communication is a central characteristic of humanity and is influenced by all aspects of our lives. It is also a multidimensional concept that we can study from many different perspectives, as illustrated in Fig. 1. The content of speech is the information that we deliver to listeners using a language. The vocal characteristics of speech are the result of our speech production system consisting of neural control and acoustics of the lungs, trachea, and larynx and the vocal tract. The prosodic elements of intonation, tone, and rhythm are often modulated by various intrinsic and environmental factors. Prosodics is often thought to be a part of paralinguistic communication, which also contain body language and other behaviors. In conversational behavior, the speech is influenced by other participants. Typical research topics in this area are conversational turn-taking and back-channeling behaviors that may reflect intrinsic mental and cognitive processes, but also habits, social conventions, and environmental constraints [11].

All aspects of speech communication illustrated in Fig. 1 and described in the following can be influenced by relevant health factors. In the following sections, the goal is to give an overview and focus on areas where the relationship has been established and where it has been demonstrated that the condition can be detected using computational speech analytics and machine learning.

At the level of the dialogue, we can also consider the roles of the participants, their goals, values, and strategies. Often a dialogue can be seen as collaborative problem solving or as a negotiation between the respondent and the caller. A dialogue may have measurable outcomes. For example, in a sales call center, negotiation with a customer can lead to conversion from a free to a paying subscriber of the service. However, often outcomes are not directly available, but may influence the customer experience and give trust in the provided service in the long term.

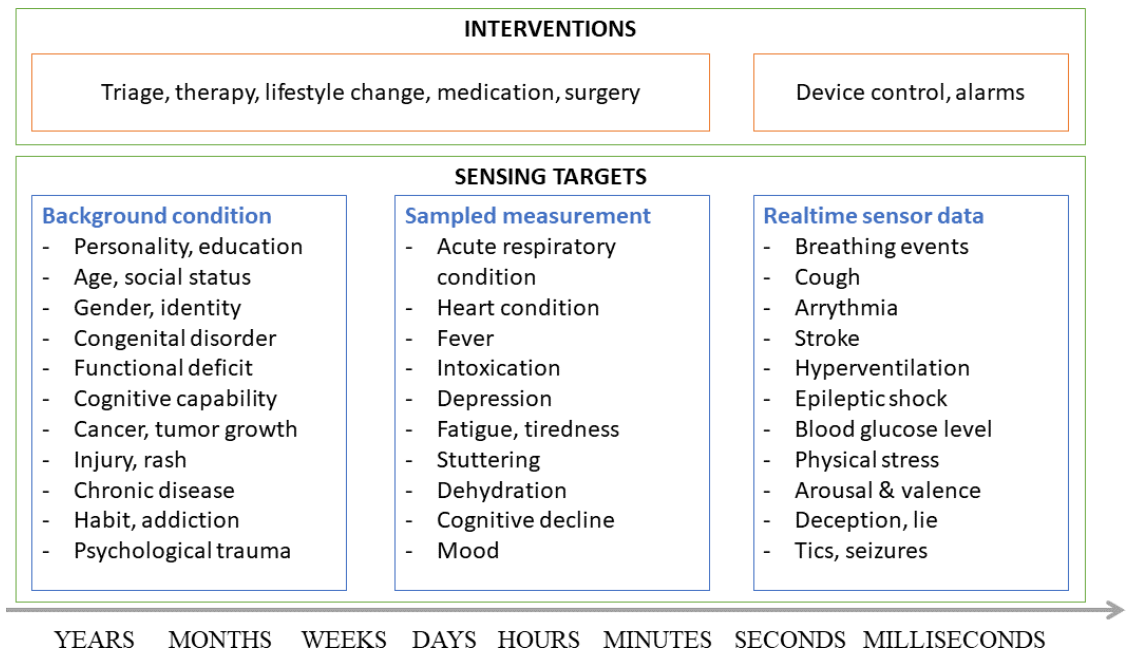

Figure 2: Sensing targets and interventions in different time-scales

Speech is a personal, learned behavior that reflects all aspects of the individual talker. Consequently, many of the sensing targets discussed in this report are available only at the individual level. For example, a mild cognitive decline or increase in the roughness of a voice can only be assessed in relation to historical data from the same subject. This may also look problematic due to natural aging and changes in the environment of the talker. However, in [12], we demonstrate that it is possible to train a multitask classification system for independent prediction of age and level of cognitive decline in a talker.

We may also look at the area of vocal biomarker research on different time scales of sensing and intervention; see Fig. 2. Longitudinal speech recordings can be used to detect background conditions and gradual changes in various factors of personality [13] and health. In a shorter time scale, speech can be used as an unobtrusive measurement for various factors that are traditionally measured by sensors attached to the body, e.g., the blood glucose level [14]. Finally, we can use speech and vocalizations for alarms and precise control of, for example, a ventilatory support device used by a patient [15].

The outline of the paper is as follows. We recapture various aspects of vocalizations with respect to speech and respiration in Section 2. The survey continues with an overview of the determination of physiological health conditions (Section 3) from vocal biomarkers followed by cognitive and mental health conditions (Section 4). Section 5 is devoted to applications, and Section 6 is a first attempt to arrive at a qualitative performance overview of various vocal biomarkers. The paper closes with a conclusion and discussion of some of the future challenges in the research of vocal biomarkers.

## 2 Vocalization aspects: speech and respiration

In this section, aspects around speech respiration are recaptured. We start by considering aspects of speech content (Section 2.1) and neural control (Section 2.1). We follow up with respiratory aspects (Section 2.3), paralinguistic (Section 2.4), and insights from conversational behavior (Section 2.5).

## 2.1   Content of speech

The analysis of transcripts of counseling sessions has a long history, especially in psychiatry [16]. Traditionally, text analysis was performed by a human expert, but a similar performance was already demonstrated for a computational method 20 years ago [17]. Recent work on text-based detection of depression in a popular data set shows an accuracy of 66% for a single conversation turn and close to 100% when accumulated over the entire conversation [18]. Performance is robust for manual and automatic transcriptions of original clean speech recordings [19]. In the last two years, we have seen rapid progress in the performance of large language models in understanding human speech [20]. This means influencing all tasks where the speech content is relevant.

Aphasia is a common neurodegenerative disorder that often follows a stroke, head injury, or a brain tumor [21]. Although the disorder primarily influences the speaker's vocabulary, it naturally has also an impact on paralinguistics, the fluency of speech production, and conversation. There are distinct types of aphasia. Fraser et al. demonstrated over 90% accuracy in detecting agrammatic aphasia, characterized by the omission of words, based on analysis of transcriptions of 300-400 word long free narratives [22].

Patients with aphasia often suffer from anomia, characterized by difficulties in naming objects. Peintner et al. [23] compared the detectability of patients with aphasia from healthy controls in an automatic detection task using pure linguistic, prosodic, or a combination of both types of features. The precision when using only linguistic or prosodic features gave an accuracy of around 80% in their data set of 3-5 minutes of speech recordings. The combination of the two feature types increased accuracy performance. A recent paper by Le et al. [24] studied the separate detectability of different paraphasias in patients with aphasia. Typical paraphasias are semantic and phonetic errors in which target words are replaced semantically or phonetically by similar words, respectively. In neologistic errors, the replacement word is not related to the original target word. The detection of these categories requires a high level of understanding of the content of speech, which was achieved in this case using conventional speech features and deep recurrent neural networks (RNNs). The best results were obtained for phonetic paraphasias, but the results are moderate. The F1-score, which is a commonly used measure to characterize the performance of a classifier, gets a value in 0.5-0.7, on a scale from 0.0 to 1.0.

The difficulties in finding words in aphasia caused by stroke (Wernicke's aphasia) or by dementia are different [25]. Patients with dementia, especially in the early stage of the disease, are less impaired in spontaneous speech. However, deterioration of lexical-semantic processing in the brain starts and becomes detectable, already in an early phase of the disease [26].

## 2.2   Neural control of speech production

Dimauro et al. [27] compared the word error rates of an ASR system in patients with Parkinson's disease (PD) and healthy controls. Word error rates were significantly higher in the PD population. In PD, the lower intelligibility of speech is not necessarily related to cognitive functions, but it reflects typical symptoms of PD that include tremors and rigidity and slow movement of the muscles, bradykinesia. The severity of PD is often characterized using the Unified Parkinson's Disease Rating Scale, UPDRS [28]. The detection target in PD speech is typically the patient's UPDRS score.

Dysarthria is a common name for speech impairments caused by difficulties in the muscular control of speech caused by a neurological problem in the motor control of speech muscles. Common conditions include Parkinson's disease, amyotrophic lateral sclerosis (ALS), and multiple sclerosis (MS) diseases. The severity of dysarthria can be described on a subjective four-level Dysarthria Severity Degree (DSD) scale. Laaridh et al. [29] have demonstrated that DSD can be predicted from a minute of text reading, but the precision is not very high. Changes in the larynx caused by rheumatoid arthritis have been shown to change the quality of patient voice [30].

Voice disorder, according to the definition of American Speech-Language-Hearing Association (ASHA), occurs when voice quality, pitch, and loudness differ or are inappropriate for an individual's age, gender, cultural background, or geographic location [29]. Typical characterizations of voice disorders are listed in Table 1. Voice disorders are categorized into Organic or Functional disorders depending on whether there is a clear physiological reason for the disorder or not, respectively. An example of a functional disorder is temporary vocal fatigue of the talker. Organic disorders may be caused by a structural or neurogenetic reason. In this terminology, dysarthria is an organic neurogenetic condition.

The detection of a dysarthric condition is often performed in a protocolled speech task. For example, in [31], dysarthric speech due to Parkinson's disease was detected in a task that consisted of fast repetition of pa/ta/ka syllables and a text reading task. In the study, the accuracy of the detection task ranged from 60 to 99%.

| Name | Description |
|---|---|
| roughness | perception of aberrant vocal fold vibration |
| breathiness | perception of audible air escape in the sound ... |
| strained quality | perception of increased effort; tense or harsh... |
| strangled quality | as if talking with breath held |
| abnormal pitch | too high, too low, pitch breaks, decreased pit... |
| abnormal loudness/volume | too high, too low, decreased range, unsteady v... |
| abnormal resonance | hypernasal, hyponasal, cul de sac resonance |
| aphonia | loss of voice |
| phonation breaks | breaks in speech voice |
| asthenia | weak voice |
| gurgly/wet sounding voice | saliva effects |
| hoarse voice | raspy, audible aperiodicity in sound |
| pulsed voice | fry register, audible creaks or pulses in sound |
| shrill voice | high, piercing sound, as if stifling a scream |
| tremulous voice | shaky voice; rhythmic pitch and loudness undul... |
| high effort | increased vocal effort associated with speaking; |
| low fatigue | decreased vocal endurance or onset of fatigue ... |
| variability | variable vocal quality throughout the day or d... |
| coughing | frequent coughing or throat clearing (may wors... |
| tensioned | excessive throat or laryngeal tension/pain/ten... |

Table 2: Voice disorders by American Speech-Language-Hearing Association (ASHA)

| Non-verbal tics | Verbal tics |
|---|---|
| Throat clearing | Repeating parts of words, or phrases |
| Sniffing | Coprolalia |
| Coughing | Prosodic changes |
| Yelling | Stuttering |
| Hiccupping | Talking to oneself (different identities) |
| Snorting | Derogatory phrases or statements |
| Animal sounds (e.g., barking) | Repetitive sounds |
| Belching | Silent palilalia |

Table 3: Typical speech effects characterized by Tourette's syndrome

Damage to muscles or other structures involved in speech production, for example, caused by an injury or a cancer, may lead to changes in articulation. In some rare cases, such as cardio-vocal syndrome, the characteristics of the voice may change due to heart disease [32] and this can be detected even in smartphone recordings [33]. Tourette's syndrome is characterized by various effects on speech motoric and language [34]. Some of the typical speech and language symptoms, see Table 3, are nonverbal motoric behaviors and some changes in prosody or linguistic content. The use of speech as a modality for triaging potential patients with TS could be particularly useful for young children [35].

## 2.3   Respiratory functions

The speech production is also influenced by infection of the lungs and respiratory tract. In a recent challenge, the task was to detect a subject who had a cold [36]. The data set had 630 talkers and the data was labeled according to the Wisconsin Upper Respiratory Symptom Survey (WURSS-24) score, which ranks the severity of the disease into seven levels ranging from not sick to severely sick [37]. The data contains scripted and free speech. The best results in the detection of speech respiratory infection had an unweighted average recall (UAR) greater than 90% [38]. UAR measure is often used to compare classifiers in speech detection tasks and represents the mean of the recall values computed for the two classes.

The detection of coughs is a widely studied subject [39] as it is a symptom of many respiratory diseases; see also Section 3.4. Cough is used in different ways. One is the detection or monitoring of cough and is a longitudinal measurement. The other is cough information extraction and is a spot check. Obviously, these can be combined in a two-step approach. Cough sounds have been studied in terms of their character (wet / dry, productive / nonproductive, voluntary / spontaneous), their source (male / female), health status (healthy / unhealthy), lung condition (obstructive / restrictive / normal) and disease diagnosis (croup, COPD, asthma, pneumonia, pertussis, tuberculosis, bronchitis, COVID-19, cold, heart failure, lower/upper respiratory tract disease). Various devices can be used, such as wearables (acceleration sensors in various areas of the body have been used), dedicated tube systems, or microphone recordings. It is generally accepted that cough is characterized by large inter-person and intra-person variability. This may be the reason why the performance reported varies quite a lot. ResAppHealth is the main proponent of disease diagnosis; see Section 3.4. Cough detection and classification has received quite a bit of attention, e.g. [40–42]. The latest deep learning algorithms achieve very good performance for the detection of single cough sounds [43, 44]. Di Perna et al. [45] reported a system for long-term, privacy-preserving, unobtrusive monitoring of cough. However, a good classifier developed in the lab does not immediately imply success for clinical applications. The gap between cough detection systems and clinical applications is still not fully bridged: What is missing are trials showing the effect of cough monitoring combined with clinical intervention; see Section 3.4.

Detection of a risk of sleep apnea from speech signals [46, 47] has also been considered. However, the results seem mixed [48]. During the severe acute respiratory syndrome coronavirus 2, SARS-CoV-2, or the COVID-19 epidemic, much attention has been paid to tracking symptoms from speech recordings. Shortness of breath, or dyspnea, is one of the key indicators detectable in the early stages of the disease. The current authors have developed technology for deep respiratory sensing from speech [49] and the detection of mild dyspnea from pairs of speech recordings from the same individual [50] and describe this work in more detail in Section 4.6. Regarding COVID-19, shortness of breath and self-reported hoarse voice have also been shown to be among the most important symptoms predicting the long condition of COVID-19 [51] where symptoms often continue for several months after the initial infection. A recent study using convolutional networks for the classification of COVID-19 from crowd-sourced speech and cough recordings demonstrated convincing results with AUC of 0.846 [52]. The paper also gives a good overview of various recent COVID-19 speech data collection studies.

## 2.4   Paralinguistics

Complex cognitive functions related to affect and emotion can influence speech communication at all levels of Fig. 1, see [53] for an overview. In particular, the detection of depression [54] and bipolar disorder [55] has been a popular research topic in the areas of speech, language, and Human-Computer Interaction (HCI) studies. A recent review by Morales et al. [56] aimed to compare the results of the literature on the detection of depression from different detection modalities, including speech, text analysis, and visual communication data. The results were not conclusive because of incomparability. Typical accuracy values ranged from 0.6 to 0.85, and the authors gave a general recommendation to use multimodality features in the detection task. In an overview of the detectability of depression and suicide risk, Cummins et al. [57] concluded that state-of-the-art methods for depression detection reach an accuracy of 79% .

Detection of distress or arousal in crisis hotline services [58] and other medical call center services [59] is an important application for speech analysis. Many commercial call center systems also provide detection of arousal based on voice and sentiment analysis of transcribed speech. Although dementia has less impact on speech fluency, mild dementia can already lead to changes in paralinguistics that can be detected from the speech signal. For example, a study showed 85% precision in the detection of mild dementia based on prosodic characteristics [60]. In our recent work that combined prosodic and linguistic for the detection of dementia condition, we obtained slightly higher performance [61].

The duration of pauses during speech in patients with ALS has been shown to differ significantly from healthy controls [62]. Changes in speech rate are also characteristic of various forms of aphasia such as primary progressive aphasia [63]. The articulation rate can be reduced from an average of 5 syllables per second in healthy controls to 1-2 syllables per second. The duration of pauses also increases in the same proportion.

## 2.5   Conversational behavior

Typical phenomena in conversational behavior are turn-taking organization, backchanneling, accommodation, and response times. Turn-taking organization refers to the way the target talker takes a turn in the conversation. Backchanneling contains, for example, verbal and non-verbal acknowledgments during the speech of the other participant. Accommodation refers to various ways on how a talker adapts the vocabulary and speaking style to the other. These phenomena are naturally influenced by various health, social, and environmental factors, see [64]. An example of accommodation is entrainment in which a talker adopts the speech rhythm of another talker. There is evidence that entraining is less accurate in Parkinson's patients than in healthy controls [65]. Otherwise, it seems that the area of automatic conversation analytics for health conditions is not widely explored. So far, research has focused mainly on applications in psychiatry, certain brain diseases, and developmental problems in children. Clinical interviewing is a popular way to diagnose various health conditions [66]. Interview techniques such as motivational interviewing [67] or Cognitive Behavioral Therapy [68] in psychology, or Clinical Dementia Rating [69] or psycholinguistic assessment for aphasia [70] have been shown to be effective as diagnostic methods or interventions. The interview protocols mostly focus on the content, but practitioners naturally also use non-verbal cues to guide the interviewing. The use of automatic speech analytics has often been proposed as an additional tool to help in clinical interviews, yet the author of this report is unaware of established protocols for it. Scherer et al. [71] studied speech changes in a conversation between an interviewer and patients with different severity of depression. Their results showed that the accommodation measured by the correlation between the patient's speech and the practitioner was reduced in more depressed patients. A high level of accommodation is often suggested to be related to a higher positive attitude and liking, and therefore the result is understandable [72]. Stolar et al. [73] also concluded that there was a clear change in the speech parameters of the interviewer, i.e., more breathy voice, with mode depressed patients. They reported a 94% classification accuracy in the detection of a depressed or non-depressed adolescent in conversation dynamics in parent-adolescent conversations, but the data set was small.

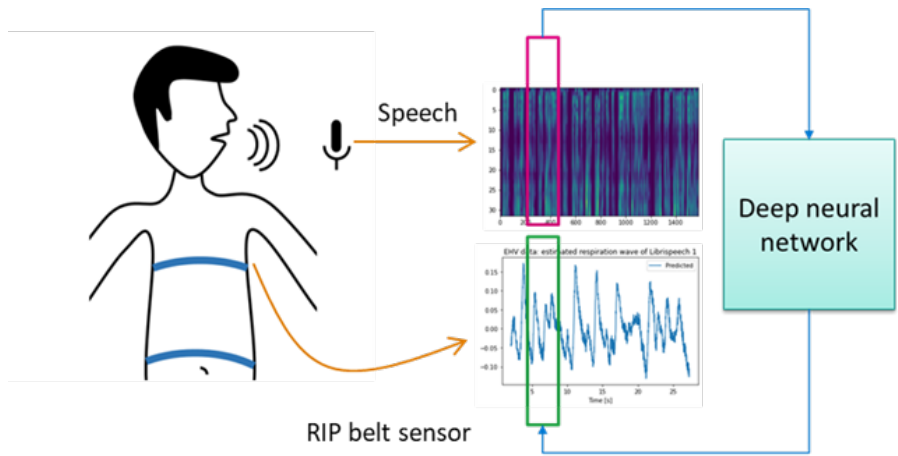

Figure 3: Virtual Respiratory Belt sensing method where the goal is to predict the signal of a RIP belt from normal conversational speech.

## 3 Physiological health conditions

In this overview, we make a distinction between the use of speech for physiological, cognitive, and mental conditions. Obviously, there is a lot of overlap between the domains, and we could also talk about the impact of systemic health on speech. In this section, the focus is on sensing targets that relate to the health of organs such as the lungs, heart, kidneys, or tissues forming the vocal tract.

This section is devoted to information on vocal biomarkers regarding physiological health. The first three sections are related to breathing that covers respiratory health (Section 3.1), breathing behavior (Section 3.2), and spirometry (Section 3.3). The next sections cover conditions that are regularly encountered: cough (Section 3.4), heart health (Section 3.5), blood sugar level (Section 3.6), dehydration (Section 3.7), and oncology (Section 3.8).

### 3.1 Sensing of respiratory health

Speech and respiration are closely related in the sense that speech is produced by organs evolved for respiratory function of the body. It implies that the speech production process interferes with the standard way of regular breathing. Evolution mitigated this issue by an unconscious planning process for respiration to be in place while speaking [74–76]. When the respiratory condition is impaired, it means that normal respiratory planning is shortened or does not operate, resulting in frequent breaks during an utterance due to the need of air. Measurement of respiratory phases during speech is possible with a respiration belt or an airflow sensor. However, such an approach is in almost all cases rather awkward. The Virtual Respiratory Belt (VRB) [49] offers a solution here, as these phases can be identified from speech and an estimate of the belt signal can be obtained.

## 3.2 Breathing behavior

The details of the breathing pattern during speech may be indicative of various physiological health conditions of the talker. Lung diseases are often classified as restrictive and obstructive diseases. In obstructive lung diseases, the patient has difficulties in exhaling air from the lungs. The most common causes of obstructive lung disease are:

- Chronic obstructive pulmonary disease (COPD), which includes emphysema and chronic bronchitis
- Asthma
- Bronchiectasis
- Cystic fibrosis

In restrictive lung diseases where the patient has difficulties in filling lungs with air some of the typical conditions are:

- Interstitial lung disease, such as idiopathic pulmonary fibrosis
- Sarcoidosis, an autoimmune disease
- Obesity, including obesity hypoventilation syndrome
- Scoliosis
- Neuromuscular disease, such as muscular dystrophy or amyotrophic lateral sclerosis (ALS)

Analysis of the breathing pattern and the search for deviant patterns in the inhalation/exhalation phase of the breathing pattern could help diagnose the nature of lung disease. Information about the estimated breathing pattern during speech can also be used in different ways. For example, our recent paper proposed smart control of a portable oxygen concentration device based on the forecasting of the next inspiration event from speech [15]. Respiratory behavior reflects speech planning [74, 77] and, consequently, the observed respiratory pattern will demonstrate the planned structural organization of the speech content. For example, an inspiration event is often placed at the end of the sentence, and, therefore, the detection of an inspiration event indicates the intended end of the sentence of the talker. This helps to resolve ambiguities in sentence parsing in automatic speech recognition and detect parts of audio where the subject is not talking for background noise estimation.

## 3.3 Spirometry

Spirometry is considered the gold standard for evaluating pulmonary function for diagnosis of obstructive and restrictive respiratory disorders. A typical spirometer consists of predictive equations that are calibrated according to the age, gender, ethnicity, height, and weight of the patient. The test involves a deep maximal inhalation by the patient and a rapid and forceful exhalation through a mouthpiece. The spirometer measures the volume flow of exhaled breath and estimates parameters like Forced Expiratory Volume in 1 second (FEV1), Forced Vital Capacity (FVC), FEV1/FVC, Peak Expiratory Flow (PEF), Flow-Volume Curve, Pressure-Volume Curve, and Work of Breathing. These parameters are used to assess pulmonary health of the patient and diagnose any respiratory disorders, for example, based on the shape of the flow-volume curve.

In scenarios where spirometry results are uncertain, bronchoprovocative tests are used to confirm diagnosis. For example, methacholine, which causes narrowing of the airways in the lungs, is used for the diagnosis of asthma along with spirometry [78]. As an alternative to medical-grade spirometers, which are expensive and have limited availability, portable spirometers are offered by several vendors (GoSpiro, SpiroLab, SpiroMax, etc.). These are shown to be as accurate as medical grade spirometers, e.g., in the diagnosis of COPD [79]. In the past, audio- or speech-based spirometry has been attempted. Samsung Research in the USA, in collaboration with Brigham and Women's Hospital (BWH), have conducted clinical studies to collect data from healthy subjects and patients with respiratory disorders (COPD / asthma) to build audio-based models to estimate spirometry parameters [80, 81]. The team collected data on 211 subjects from 3 studies, in which participants were asked to read the 'rainbow' paragraph loudly, and the audio was recorded using the Samsung Galaxy Note 8 smartphone mic. In addition to the read speech data, data on exhalation sounds while performing the spirometry protocol, voluntary cough sounds, and vowel sounds (Aa, . . . ) were recorded. One of their studies with BWH involved giving several doses of Methacholine to participants and recording their spirometry and audio data after each dose. They have shown that the voice loudness, pitch and pauses are correlated to the flow-volume curves obtained from spirometry and can be used to estimate parameters like FEV1, FVC, etc. They show that use of exhalation sounds, while following the spirometry protocol, results in better accuracy in estimating the spirometry parameters from audio. They also quantify the quality of the user's exhalation effort, which is informed by the guidelines of the American Thoracic Society.

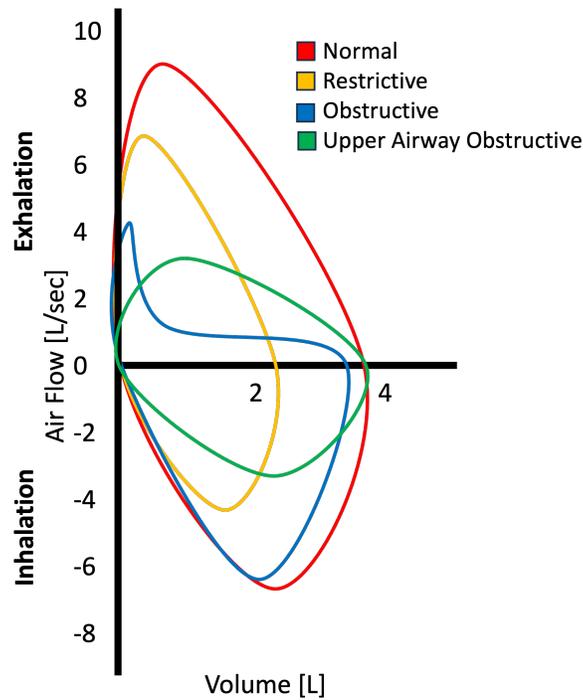

Figure 4: Flow-Volume curve from spirometry with inhalation (bottom) and exhalation (top) curves. Morphology of the curve indicate type of respiratory disorder.

Researchers at the University of Washington have done similar work showing that forced exhalation sounds from multiple microphones with compressed or uncompressed audio data can be used to estimate spirometry parameters with reasonable accuracy [82, 83]. They also propose the use of a 3D printed vortex whistle (for which pitch is proportional to airflow), specifically for patients with low lung function. A whistle-sound-based solution is shown to reduce false negatives in determining lung function and is more accurate in estimating flow-volume curves. The loss in mean performance from a call-in service (over a GSM network), compared to locally available uncompressed audio data, is also shown to be only around 1%, while estimating some of the spirometry parameters of audio.

### 3.4 Cough screening

Cough sound acquisition comes in two major variants. One is the detection or monitoring of cough and is a longitudinal measurement. The other one is information extraction from cough and is a spot check. Obviously, these can be combined in a two-step approach.

Cough sounds have been studied in terms of their character (wet / dry, productive / nonproductive, voluntary / spontaneous), their source (male / female), health status (healthy / unhealthy), lung condition (obstructive / restrictive / normal) and disease diagnosis (croup, COPD, asthma, pneumonia, pertussis, tuberculosis, bronchitis, COVID-19, cold, heart failure, lower/upper respiratory tract disease). Various devices can be used, such as wearables (acceleration sensors in various areas of the body have been used), dedicated tube systems, or microphone recordings. It is generally accepted that cough is characterized by large inter-person and intra-person variability. This may be the reason why the performance reported varies quite a lot. ResApp Health reported a quite successful performance in disease differentiation [84]); results reported by other parties are less convincing [85].

Monitoring can be done by mobile or stationary systems, on-body accelerometers, or microphones. There is a clear desire from clinicians for insight into (amount of) coughing for chronic coughers. Also, pharmaceutical companies need cough monitors for testing/confirming effects of, e.g., antitussive medication. From the perspective of a chronic patient, unobtrusiveness is desired for obvious reasons; for the caregiver, this is also a desired property, as it greatly increases adherence. Promises here are early warning systems (leading to less/shorter hospital admissions), medication control (leading to better medication usage), and early discharge by deploying convalescence monitoring. Hyfe proposed, among others, TB [86] and COVID [87] as application areas based on the results of the frequency of cough at the cohort level. Translation to a single-patient system in real-life situations (patient homes), covering aspects of proper metrics, personalization, cough prevalence, medical outcomes (e.g. exacerbation) and early warnings (lead time), are still under development. The first instance of an unobtrusive night-time cough monitor in which such aspects were considered was reported in [45, 88–90] using a set-up that allows personalization without intrusion into privacy.

## 3.5 Heart health sensing from vocalizations

Beyond Verbal, currently part of Vocalis Health, has worked with the Mayo Clinic to detect certain heart conditions from speech signals. With a cohort of 101 patients, Maor et al. [91] were able to predict the mortality of CHF patients from a 20s speech recording. Certain heart diseases may also lead to a rare condition of cardio-vocal syndrome, an audible hoarseness [92] of the voice due to a structural change in the arteries that impinges on the left recurrent laryngeal nerve that controls the muscles of the larynx [32]. Cardiokol has technology and an app to detect arrythmias in sustained vowels. In their 2021 paper [93] the accuracy in the detection of arterial fibrillation, AF, from a sustained 'Ahh' vowel was reported in a group of 143 patients 95%. Gouda et al. [94] have proposed speech sensing and voice assistants as useful tools for the diagnosis and management of the health lifestyle of cardiac patients.

## 3.6 Blood sugar level

The sensing of blood glucose level is very important especially for Type I diabetics and the possibility of using speech has been considered a very attractive way of monitoring the glucose level without being obtrusive. Some studies, see [14], report the accuracy of 79% in the detection of different levels of blood glucose concentration from short speech utterances collected using a smartphone and a continuous glucose monitoring device. In a meta-study [95] it was concluded that detection appears to be possible at the individual level, but interpretation of the results is often problematic, for example, because the subjects are aware of their blood glucose level. It is usually thought that there are two mechanisms in the blood glucose level that can influence speech:

- Chance in the blood flowing in the larynx and the glottis causes changes in the elastic properties of the organs which would change the glottis in voiced speech.

- The mood changes related to hypoglycaemia, feeling of anxiety, and hyperglycaemia, which may cause arousal, respectively, are reflected in changes in the prosody and content of speech.

## 3.7 Dehydration and fluid accumulation

The body fluid balance may have various effects on speech. Speech itself is known to cause dehydration of the vocal cords, which can cause hoarseness in the voice after long speech [96]. Dehydration increases the respiratory rate. Systemic dehydration also influences speech in a similar way [97]. Dryness of the tongue, lips, and other tissues of the vocal tract may be due to systemic dehydration of the body. The authors of this paper are unaware of the literature on the use of speech audio for the detection or quantification of the level of dehydration. However, drug-induced dehydration has been shown, for example, to increase the phonation threshold pressure, PTP, that is, the subglottal pressure required to start vocal utterances [98]. The indirect estimation of PTP from speech would be a potentially interesting detection target, for example, in the evaluation of adverse medication side effects [99].

## 3.8 Oncology

Lung cancer is the type of cancer that is the largest in terms of burden and deaths globally and is typically diagnosed with a CT scan. It is known that in certain populations, COPD (detected by spirometry) can be a precursor to lung cancer, and its sub-typing can be used for effective lung cancer surveillance [100]. In addition, respiration patterns show changes in patients undergoing radiation therapy [101] and chemotherapy [102]. Some of these lung functions can be determined with high precision using spirometry. Vocal biomarker technology can be integrated in any speech interface, for example, in conversational health counseling services for cancer patients [103].

As described in Section 3.3, virtual spirometry can be performed using audio or speech data with good accuracy. Audio/speech data from different protocols (paragraph read, exhalation, vowel and cough sounds), patient information (age, gender, height, weight), and other vitals (HR, SpO2, HRV) can be used as inputs to AI algorithms to predict RIP and spirometry data. Direct estimation of spirometry parameters using speech can potentially offer a safer (radiation-free), accessible, and cheaper option to detect and stage lung cancer patients. Also, patients with similar audio signatures can be matched to clinical trials conducted for lung cancer treatment. The tracking of changes in audio can enable the monitoring and clinical surveillance of patients undergoing radiotherapy or chemotherapy.

The cancers in the area of the vocal tract, including trachea, larynx, and mouth, have effect on voice of the talker. Halpern et al. [104] reported 88% accuracy in the detection of speech of an oral cancer patient from a healthy control.

Brain cancers may lead to changes in cognitive and motor functions that may affect voice and speech content.

# 4 Cognitive and mental health conditions

This section is devoted to information on vocal biomarkers on cognitive and mental health. The first sections are related to three diseases: Alzheimer (Section 4.1), Parkinson (Section 4.2) and autism (Section 4.3). Section 4.4 considers emotion and sentiments. The following sections address the states of depression (Section 4.5), intoxication (Section 4.6) and fatigue and sleepiness (Section 4.7).

## 4.1 Alzheimer's disease

The detection of AD from speech was already discussed in Chapter 2. According to WHO, about 55M people globally suffer from dementia and about 60- 70% of them probably have AD. The expected global market size in 2029 is $6.3B and a large part of it is in medication, although, there is no drug to cure the condition. Many patients are undiagnosed and the progression of their disease could be potentially slowed down by nutrition, lifestyle changes, and medications. Therefore, early detection is considered particularly important.

## 4.2 Parkinson's disease

The expected market size of Parkinson's disease, PD, is also increasing rapidly and is expected to reach globally $6.7B by 2030. About 70% of the market is the cost of medications, and the rest, various therapies, and care settings. Medications are known to be helpful in reducing symptoms and slowing the progression of the disease, but common medications have significant side effects. Also in PD, for the same reason as in AD, early detection and various therapies in the early stage of the disease would significantly reduce costs and improve the quality of life of patients.

## 4.3 Autism spectrum disorder

Autism is frequently accompanied by communication difficulties, including delayed speech development, difficulties in language comprehension, and social communication challenges. The size of the market for early detection and treatment is expected to reach the size of the market $5 B in 2030. Diagnostic evaluations commonly deploy speech assessments that measure speech production (e.g., articulation, fluency), language comprehension, and social communication skills (e.g., initiating and maintaining conversation, interpreting non-verbal cues). Assessments of pragmatic language, which capture the use of language in social contexts, can further assist in the identification of autism. Given their efficacy in detecting autism, speech and language assessments constitute valuable tools in the diagnostic process. Several studies have reported that children with autism exhibit distinctive breathing patterns, which can include slower or shallower breathing or differences in respiratory rate variability [105]. These deviations from the breathing pattern may be linked to other autism symptoms, such as sensory processing difficulties or anxiety. Specifically, children with autism have been observed to have lower respiratory sinus arrhythmia (RSA) values, indicating atypical regulation of the autonomic nervous system. Furthermore, compared to typically developing children, children with autism exhibit reduced respiratory variability, which can imply a decreased flexibility in respiratory control. Additionally, children with autism may demonstrate lower respiratory sinus arrhythmia and a flatter breathing pattern, indicating difficulties in emotional regulation and social participation [106, 107]. Taken together, these findings suggest that breathing pattern assessments may be useful in identifying and characterizing autism. McDaniel et al. [108] provide a recent overview of the measurement of preverbal vocalization in infants with Autism Spectrum Disorder (ASD). Preverbal means here the phase in development where the child has no, or few, words. Some of the typical differences between typical development and children with ASD are related to the lower rate of intentional vocal communication, smaller consonant inventories, atypical vocalizations, and decreased responsivity and attention to the vocalizations of others. The detection of ASD from child vocalization has been studied using the De-Enigma data set. Typical performance for the detection of an ASD is around 70% [109, 110].

## 4.4 Emotion, state, and sentiment sensing

Emotion sensing from speech is a very popular area of research [53] where most applications are related to the detection of caller emotion and arousal in call centers and public spaces, for example. The detection of various mental diseases has been studied a lot, but the performance is usually low. For example, a recent article that uses speech data for the detection of various psychiatric conditions [54] reported a low $R^2$ value, i.e., in less than half of the cases the condition was detectable by speech data. The problems stem already from vague definitions of conditions such as schizophrenia, depression, or general anxiety, and their volatility.

## 4.5 Depression

Depression is globally a large problem that causes high healthcare care costs for societies. The global market for anxiety disorders and depression was estimated to be around $12B in 2019 and is expected to reach $16 B by 2027. Approximately one half of the market is in medication and the rest for various therapies. Speech-based sensing can be used to determine and monitor changes in the condition of the patient, for example, in response to medication. Speech-based interfaces can also be used to deliver therapeutic interventions to patients. Dyspnea can be associated with anxiety and panic, which are often comorbid with depression. Chronic diseases such as heart failure or chronic obstructive pulmonary disease (COPD) can lead to dyspnea and depression due to their impact on physical health and quality of life. People experiencing depression may have changes in their physical health, including fatigue and altered respiratory patterns, which can contribute to a feeling of dyspnea.

## 4.6 Intoxication

Intoxicated speech detection challenge was organized as a part of the Interspeech conference in 2011. The winning team received a 73% precision in detecting a sober or alcohol intoxicated (> 0.5mg/L) speaker from a single speech utterance.

## 4.7 Fatigue and sleepiness

It is generally assumed that fatigue and sleepiness affect the character of the voice. For example, it has been demonstrated that the sleepiness level of a subject can be detected from a single sustained vowel [111] with reasonable accuracy.

# 5 Applications

An overview of the various applications that were found in the survey are structured in 13 sections running well-known tests to applications in respiratory support and speech therapies.

## 5.1 Active monitoring or protocolled speech-based health tests

There is a wide range of speech-based health tests that are used regularly in clinical practice. Many of the tests that are currently performed in a clinic can be potentially automated and turned into a home test that can be expected to improve patient experience, health outcomes, and cost of care. Automation of routine testing also saves time for healthcare professionals.

## 5.2 Diadochokinetic speech tests

Diadochokinetic (DDK) tests are regularly used to assess neurological speech motor impairments [112]. There are two common variants of the test: the alternating motion rate (AMR) test and the sequential motion rate (SMR) task. In AMR the subjects repeat as quickly as possible sequence of identical syllables, e.g., "pa-pa-pa-.." or "ta-ta-ta-. . ." and in SMR variant sequences of syllables, that is, "pa-ta-ka-pa-ta-. . .". Conventionally, recordings are analyzed manually, but automated techniques based on deep learning have recently been demonstrated [113].

## 5.3 Cognitive retrieval tests

Boston Naming Test (BNT) is a widely used test for children and adults with aphasia [114]. In the test, the patient is requested to name entities shown in 60 drawings which are ordered by difficulty. The examiner scores each answer based on the responses of the patient. BNT test has been criticized for being culturally biased and lacking in accuracy [115] and proposed to be replaced by a more comprehensive test such as the Neuropsychological Assessment Battery, NAB [116]. NAB has 33 different tests for cognitive capability, learning, and memory. Montreal Cognitive Assessment, MoCA [117], is a tool for assessing mild cognitive impairment and contains a memory recall test and speech fluency tests. In typical speech fluency tests, the patient is asked to list, as fast as possible, words starting with a particular letter, or names of entities from a category, e.g., names of animals. Fluency tests have been validated in many studies; see, e.g., [118]. Controlled Oral Word Association Test, COWAT [119] is a variant of the verbal fluency test where the subject lists, in one minute, words starting with one letter and then repeats the task with two other letters.

## 5.4 Tests on vocal development in children

There are several tests that help to assess child development. Early Communication Indicator, ECI [120], is a test for toddlers in which the parent plays with a particular toy with the child and communicates with the child. The score is based on the count of various kinds of vocalizations and gestures in a six-minute test session. Autism Diagnostic Interview (ADI-R) is a structured interview with a physician and a parent of a child with 93 questions to help distinguish ASD from other developmental disorders [121]. The interview and scoring of the results require extensive training and the interview takes approximately two hours.

## 5.5 Passive Monitoring

The main use case considered in this report can be called passive monitoring of the health condition of a patient. The sensing can be performed during any calls with the subscriber and use the natural speech content of the call. It is also possible to set up calls for speech monitoring and have protocolled conversations with the patient, which could be considered as an example of active monitoring. Active speech-based detection is common practice in many brain diseases and is also the basis for the diagnosis of, for example, aphasias.

Typically, the results of passive speech monitoring could be given to caregivers in a monitoring dashboard. For example, this information may show some trends in the development of the cognitive state, communication capabilities, or respiratory functions of the patient. The caregiver can use the information to plan the care of the patient. The derived health information can also be used as features for risk prediction.

## 5.6 Assisting technologies for call centers

The information conveyed in speech may be directly related to the health of a person. For example, if a person says that she has fallen or has chest pain, this information may be extracted from the speech using Automatic Speech Recognition (ASR) and Natural Language Understanding (NLU) technologies. A typical task in NLU would be to detect that a patient reports a certain health condition, for example, "I have breathing difficulties." Using vocal biomarkers, the respiratory health condition can be detected directly from the way the person is breathing or talking during the conversation. In the final application, of course, both NLU and speech-based sensing can be used. NLU technology and speech analytics, in general, can be, for example, used for automatic generation of case notes from calls. Automatic note generation may reduce the time of a call significantly. Optimization of the average speed of answer and handling times in call centers, or contact centers [122], by using computational techniques is an active area of research; see, e.g. [123].

In recent years, we have seen significant developments in ASR performance [124], which has led to the introduction of spoken dialogue systems for home use. Smart assistants such as Amazon Alexa, Google Home, and various others already offer various home health services that help monitor and find solutions to health problems.

## 5.7 Automation of call center

The development of assist technology for better logging of call outcomes can be seen as the first step toward partial automation of the call center services [125] with the help of automated agents. In the first phase, automation can be introduced as an assistive technology that would give the PRA reminders on what to ask and how to run the conversation in the most effective way for the given customer. A large number of PERS calls are not emergency calls, but relatively simple technical questions, for example related to battery changes. Those calls could be gradually replaced by automated call agents. The use of automated conversational agents for outbound calls is also being introduced, for example, this is one of the proposed use cases for Google Duplex.

The development of automated health counseling systems [126] based on conversational interfaces is an area that grows very rapidly. The effectiveness of automated agents in depression counseling has recently been demonstrated in a random controlled trial [127]. The results of chatbot interventions in clinical psychiatry are still inconclusive; see, e.g. [128, 129], one can anticipate their use in contact center applications in the future. A recent meta-analysis [130] concluded that chatbots are already effective for dietary, physical activity, and sleep counseling.

### 5.8 Home healthcare and tele-health

Speech-based sensing makes it possible to monitor the state and changes in the patient's condition. Listening to a patient for health monitoring purposes is nothing new in health care. In many home healthcare applications and services, communication between caregivers and patient takes place over a telephone or video conference link. The caregiver gets information about the patient's health status from the patient's responses and from various nonverbal cues. In particular, the cognitive and mental state or, for example, the respiratory conditions of a patient may lead to changes in spoken content, intonation, voice, and non-verbal conversational behavior [131]. When the caregiver knows the patient well, the monitoring of speech and nonverbal cues in conversation is a natural and instinctive part of the care process and does not necessarily require a technological solution based on speech analytics. In a call center where each respondent may have thousands of patients, this is more difficult, and in a case where the interaction takes place with a spoken dialogue system, the sensing can only be performed using speech technology.

### 5.9 Elderly care and chronic disease management

The elderly care market size in the US alone is estimated to be around $400 B and is expected to grow globally to $2.4T by 2028. The market consists of medications, devices, and services. Vocal biomarker technology has many potential use cases in elderly care. A large part of the market is in elderly care facilities where passive and active speech sensing, discussed in this report, can be integrated into infrastructure and care services for residents. In addition to screening and longitudinal tracking health conditions of individual patients, passive sensing of vocalizations, for example coughing and sneezing, can be used for detecting and monitoring respiratory infections spreading in the facility.

In chronic disease management, patient education, screenings, checks, monitoring, and care coordination are expected to have a market size of $10B by 2028. This may consist of telehealth services, speech assistants in home, phone, and car applications where the availability of speech content opens possibilities for active and passive conversational health sensing.

Medication has various effects on speech and, therefore, it is possible to use vocal biomarkers to track the effectiveness of a drug to the health of the patient and detect possible medication side-effects such as fatigue or cognitive and mental effects.

In tele-health services and Personal Emergency Response Systems (PERS) for home dwelling frail elderly speech information is naturally available for speech-based sensing applications. In our analysis of the former Philips Lifeline data it was found that a large number of PERS patients with a pendant and a speakerphone were in contact with the call center very regularly, daily or even multiple times a day.

### 5.10 Cardiac health

The respiratory belt signal measures the expansion and contraction of the chest as a result of breathing. The ECG signal, on the other hand, measures the electrical activity of the heart. The ECG signal provides information about the rhythm and function of the heart, while the respiratory belt signal provides information about the rate and depth of breathing. The relationship between these two signals is that changes in respiratory rate and depth can affect heart rate and rhythm. For example, when we inhale, the heart rate typically slows down, and when we exhale, the heart rate typically speeds up. This phenomenon is known as respiratory sinus arrhythmia (RSA) and is a normal physiological response. By analyzing how changes in respiratory activity affect heart rate and rhythm, we can gain a better understanding of how the cardiovascular system functions. This information can be used in the diagnosis and treatment of various cardiovascular conditions.

### 5.11 Respiratory support

Many lung disease patients need additional oxygen support, often based on an oxygen concentrator device that delivers oxygen-rich air to the patient's nose through a nasal cannula [132]. The POCs typically work either in a continuous flow mode or in pulse dose mode where the air flow pulse, the bolus, is delivered during the inhalation phase triggered by a pressure sensor. While the cannula allows the patient to talk freely, the flow of oxygen during speech is disturbed in both POC modes. In nonspeech breathing during light activity, the average inspiration and expiration duration in COPD patients are 1s and 1.7s, respectively [133]. During speech, the average respiratory rate is approximately half that of nonspeech breathing and the duration of inhalation is significantly reduced [49, 77, 134]. In a continuous-flow model only a fraction of oxygen gets into lungs during normal speech. In pulse dose mode, the latency of air delivery in most devices is 100-200ms and the duration of the bolus is 200-800ms [135]. Moreover, the pulse dose mode is activated by nasal inhalation while it is common during speech to inhale through the mouth, which would not trigger the pressure sensor. In [136] it was demonstrated that it is possible to forecast the inhalation event and therefore control the oxygen administration so that it is aligned with the brief inspiration events in continuous speech.

The forecasting of inhale moments from speech can be useful in various applications including open types of air masks for scuba divers, firemen, or astronauts, who need to talk naturally while wearing the mask for oxygen support, and for breathing support devices for lung patients. The professional safety mask market is expected to reach $7.6B by 2026, the medical mask market $35B in 2030, and the size of the scuba diving equipment market size was already $1.95B in 2019. Moreover, it has very interesting opportunities to enable more natural communication in various metaverse applications, which has an expected value of tens of trillions of dollars [137].

## 5.12 Speech Therapies

Speech sensing can be used to support speech therapy techniques. Speech Entrainment for Aphasia Recovery (SpARc) is a therapeutic technique used in stroke rehabilitation of patients with speech disorders [138]. In SpARc, the patient is asked to mimic the speech of talker presented by video or audio. The performance of the subject can be assessed by the percentage of correctly repeated words and by the change in various aphasia tests, such as the Western Aphasia Battery [139] or the Boston Naming Test [114].

## 5.13 Automotive applications

The size of the driver health monitoring market size is around $1 B and is expected to increase to $2.4B by 2027, in some estimates reaching globally $25B by 2030 [140]. The main driver is the fact that many accidents occur due to fatigue and sudden incidents such as cardiac arrest of the driver. In an automotive environment, speech-based sensing can be used to sense various driver and passenger conditions. Intoxication and fatigue are useful in all scenarios. Emotions such as anger or arousal, for example, can be part of safety monitoring in public transportation systems. The detection of driver health can use various health detection techniques discussed in this report. Speech-based sensing has significant advantages over several other sensing modalities. Sensors attached to the body are difficult to use in a car environment, and inertial sensors mounted to the structure of the car suffer from a high level of vibration. Cameras and IR systems suffer from constantly changing lighting conditions.

# 6 Performance of vocal biomarkers

In this review, the objective was to give a broad overview of a range of vocal biomarkers proposed in the literature. In Section 1 we define a vocal biomarker as a method to detect a condition of the talker, or estimate the measurement values of another sensory modality. The detection of a specific characteristic of a vocalization, such as a cough or roughness of the voice, is subject to acoustic measurement error. The detection of cough is a *primary measurement* of the condition for the root factor influencing of the patient. As illustrated in Fig. 5 there is often another primary measure for the underlying root factor, for example, a respiratory rate of a patient, or the Hamilton Depression Rating Scale (HDRS) for the diagnosis of clinical depression [141]. The primary measure has some error which may be caused by sensor noise or ambiguities in the rating protocol. A vocal biomarker is typically developed and validated so that it represents a translation of the primary measure. The translation is subject to a translation error. Obviously, the acoustic measurement of the biomarker is influenced by noise and distortion.

Table 4 gives a qualitative overview of different targets for vocal markers and the impact of measurement noise, or lack of the reliability of the primary measure, the anticipated translation noise from the primary measure to the vocal biomarker, and the probability of the impact of acoustic error caused by the expected low level of the acoustic signature. In most cases the acoustic error can be expected to be low because the condition influences the content and prosody of speech. The translation error is often ranked medium to high because the acoustic measure is a very indirect indicator of the primary measure. Finally, the measurement error of the primary measure is in some cases a direct indication of the condition, for example, respiratory rate and the RIP belt measurement, where we can expect that the noise influencing the measurement is generally low. In cases where the primary measure is a holistic score, for example, Montreal Cognitive Assessment, MoCA [117] based on subjective assessment of the responses of a patient, the measure error can often be considered medium or high due to low reliability and repeatability.

One could conclude that the vocal biomarkers for the detection targets, or root factors, that have at least one high noise source are probably very challenging to develop into a useful and versatile measure. The easiest cases seem to be basic respiratory measurements, infection detection, intoxication, and fatigue detection. However, it should be noted that the progress in the area of vocal biomarkers is fast and we can expect to see significant progress in the coming years in the development of new algorithms and primary measure targets.

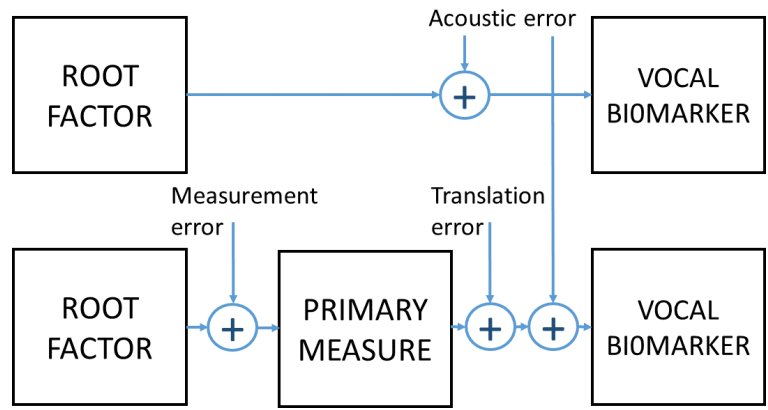

Figure 5: Vocal biomarker may represent a specific root factor, but often it is a secondary measure approximating a more intrusive primary measurement of the underlying root factor.

| Class | Root factor | Primary measure | Meas. error | Trans. Error | Acoust. Error |
|---|---|---|---|---|---|
| Cardiopulm. | Respiratory rate | RIP/Air flow | Low [142] | Low | Low |
| Cardiopulm. | Dyspnea | Spirometry | Medium | Medium | Low |
| Cardiopulm. | COPD/asthma | Clinician | Medium | Medium | Low |
| Cardiopulm. | Breathing sounds | Stethoscope | Medium | Medium | High |
| Cardiopulm. | Cough | Microphone | N/A | N/A | Medium |
| Cardiopulm. | Arrythmias | ECG | Low | High | High |
| Cardiopulm. | cardio-vocal syndrome | Imaging study | Medium | High | High |
| Infection | Flu/cold | WURSS | Low [37] | Low | Medium |
| Infection | Covid-19 | PCR test | Low [143] | Medium | Medium |
| Mental | Depression | HDRS | Medium [141] | Medium | Low |
| Mental | Schizotype | DSM-5 | High | High | Low |
| Cognitive | Aphasia | WAB | Medium [144] | Medium | Low |
| Cognitive | Parkinson's Disease | UPDRS | Medium [28] | Medium | Low |
| Cognitive | Dysarthria | DSD | Medium [112] | Medium | Low |
| Cognitive | Tourette's syndrome | Clinician | Medium [145] | High | Low |
| Cognitive | Dementia | CDR | Medium | High | Low |
| Cognitive | ALS | EMG, US | Low | High | Low |
| Cognitive | Autism Spectrum Disorder | ASQ | Medium | High | Low |
| Systemic | Intoxication | Blood: alcohol/drug | Low | Medium | Low |
| Systemic | Blood Glucose Level | Blood: sugar level | Low | High | High |
| Systemic | Systemic dehydration | Blood: serum osmolality | Low | High | High |
| Systemic | Fatigue/sleepiness | Self-reporting | Medium | Medium | Low |
| Oncology | Oral cancer | Imaging, pathology | low | High | Medium |

Table 4: Characterization of noise sources in vocal biomarkers

# 7 Conclusions and discussion

This paper aims at giving a broad and updated survey on the recent progress in vocal biomarker technology and its applications. Over the past few years, the rapid advancement of machine learning techniques has expanded the horizons of potential applications and significantly enhanced the performance in conventional problem domains. However, it is important to note that the performance of vocal biomarkers still falls short of the standards required for their use in healthcare applications. A basic qualitative overview of noise sources to determine the accuracy has been presented in order to identify promising directions.

Two challenges in this research area are particularly worth mentioning here. One challenge is that a vocal biomarker, due to training data, typically serves as secondary approximations of primary measures. In many cases, machine learning models for vocal biomarkers are trained using the primary measure rather than the ground truth. This approach introduces measurement errors into the primary measure, complicates the interpretation of vocal biomarkers in relation to the primary measure, and is susceptible to acoustic noise and distortion, which can adversely affect the performance of vocal biomarkers.

Another challenge relates to the use of vocal biomarker technology in practical applications. Various biomarkers (e.g., protocolled speech, features from sustained vowels) require a triage set-up realizable as apps or websites that collect recordings and may require trending over longer periods of time. As such, it requires (prolonged) adherence from the user or patient.